\documentclass[twocolumn]{openjournal}

\usepackage{xcolor}
\usepackage{textgreek}
\usepackage[utf8]{inputenc}
\usepackage[english]{babel}

\usepackage{hyperref}
\hypersetup{
    unicode, 
    colorlinks=true,
    linkcolor=linkcolor,
    citecolor=linkcolor,
    filecolor=linkcolor,
    urlcolor=linkcolor,
}
\usepackage{color,colortbl}
\definecolor{linkcolor}{rgb}{0.0,0.3,0.5}
\usepackage{tensind}
\tensordelimiter{?}
\DeclareGraphicsExtensions{.bmp,.png,.jpg,.pdf}
\usepackage{verbatim}
\usepackage[normalem]{ulem}
\usepackage{orcidlink}
\usepackage{soul}
\usepackage{amsmath}

\graphicspath{ {./figs/} }

\addto\extrasenglish{%

\newcommand{\aref}[1]{\hyperref[#1]{Appendix~\autoref{#1}}}
}

\newcommand{\reply}[1]{{{{#1}}}}

\begin{document}
\title{Differential virial analysis:\\a new technique to determine the dynamical state of molecular clouds}

\author{Mark R. Krumholz\orcidlink{0000-0003-3893-854X}}
\email{mark.krumholz@anu.edu.au}
\affiliation{Research School of Astronomy \& Astrophysics, Australian National University, 233 Mt Stromlo Rd, Stromlo, ACT 2611, Australia}

\author{Charles J. Lada\orcidlink{0000-0002-4658-7017}}
\email{clada@cfa.harvard.edu}
\affiliation{Center for Astrophysics, Harvard \& Smithsonian, 60 Garden St, MS 72, Cambridge, MA 02138, USA}

\author{Jan Forbrich\orcidlink{0000-0001-8694-4966}}
\email{j.forbrich@herts.ac.uk}
\affiliation{Centre for Astrophysics Research, University of Hertfordshire, College Lane, Hatfield AL10 9AB, UK}

\begin{abstract}
   Since molecular clouds form stars, at least some parts of them must be in a state of collapse. However, there is a long-standing debate as to whether that collapse is \textit{local}, involving only a small fraction of the cloud mass, or \textit{global}, with most mass in a state of collapse up to the moment when it is dispersed by stellar feedback. In principle it is possible to distinguish these possibilities from clouds' virial ratios, which should be a factor of two larger for collapse than for equilibrium, but systematic uncertainties have thus far prevented such measurements. Here we propose a new analysis method to overcome this limitation: while the \textit{absolute} value of a cloud's virial ratio is too uncertain to distinguish global from local collapse, the \textit{differential} change in virial ratio as a function of surface density is also diagnostic of clouds' dynamical state, and can be measured with far fewer systematic uncertainties. We demonstrate the basic principles of the method using simple analytic models of supported and collapsing clouds, validate it from full 3D simulations, and discuss possible challenges in applying the method to real data. We then provide a preliminary application of the technique to recent observations of the molecular clouds in Andromeda, showing that most of them are inconsistent with being in a state of global collapse.
\end{abstract}

\begin{keywords}
    {Giant molecular clouds (653), star formation (1569), astrophysical fluid dynamics (101)}
\end{keywords}

\maketitle

\section{Introduction}
\label{sec:intro}

Ever since their discovery in early CO line surveys \citep{Lada76a, Blair78a, Blitz80b}, the dynamical state of giant molecular clouds (GMCs) has been a topic of significant interest. They are observed to have large linewidths that imply the presence of significant non-thermal motions, and, especially given the strong observed correlation between GMCs and star formation \citep[e.g.,][]{Mooney88a}, it was natural to hypothesize that such motions could be indicative of large-scale gravitational collapse \citep{Goldreich74a}. However, it rapidly became apparent that this picture faced an accounting problem: given the total mass of giant molecular clouds in the Milky Way, collapse of these structures into stars on anything like a free-fall timescale would imply a star formation rate far higher than what is actually observed \citep{Zuckerman74a}.

One approach to resolving this accounting problem is to abandon the hypothesis that GMCs are in a state of gravitational collapse, and instead attribute the large observed velocity dispersions to turbulent motions that mostly inhibit collapse and reduce star formation \citep[e.g.,][]{Mac-Low04a, Krumholz05c, Federrath12a}. In this view, while GMCs are perhaps gravitationally bound on their largest scales (though this need not be the case; e.g.,~\citealt{Evans22a}), most of the mass on smaller scales within them is unbound, and at any given time only a small fraction of a GMC's mass is collapsing. The alternative approach is to retain the collapse hypothesis, and instead attribute the low galaxy-averaged star formation rate to some combination of most GMCs being unbound and non-star-forming, and to feedback being so efficient that, even though GMCs are in a state of global collapse, the tiny fraction of the mass that forms stars is able to disperse and dissociate the remaining collapsing mass \citep[e.g.,][]{Ballesteros-Paredes11a, Vazquez-Semadeni19a}. We can roughly categorize these theoretical pictures as \textit{local} collapse versus \textit{global} collapse -- see \citet{Dobbs14a} and \citet{Chevance23a} for a thorough review.

Most attempts to distinguish between these two scenarios using observations have relied either on morphological analyses of individual targets \citep[e.g.,][]{Anderson21b} or on statistical analyses of GMC populations and of the denser gas tracers within them (e.g., \citealt{Krumholz07e, Krumholz12a, Evans14a, Heyer16a, Krumholz19a, Chevance20a, Pokhrel21a, Hu22a} versus \citealt{Vazquez-Semadeni19a, Ballesteros-Paredes24a, Zamora-Aviles24a}). The former approach suffers from the difficulty that it is not quantitative and there is no clear mapping between either theoretical picture and morphology, while the latter is useful only for statistical averages, and does not provide constraints on the dynamical state of any particular cloud.

In principle if we could measure the energy budgets of GMCs with sufficient precision, this \reply{might} provide a mechanism to distinguish local from global collapse within a single cloud. The reason is that the two scenarios suggest slightly different ratios of kinetic to gravitational energy. For example, \citet{Vazquez-Semadeni07a} simulate the formation of a molecular cloud via a colliding flow, which then goes into a global collapse; in their simulations (see their Figure 8) there is a brief period of $\sim 5$ Myr when molecular material is still assembling when the kinetic energy is larger than the potential energy, but thereafter, for the remaining $\sim 30$ Myr of the simulation, the cloud's kinetic and potential energies closely track one another, differing by less than a factor of two until late times when feedback begins to disperse the gas. This is not surprising: in a system where the motions are primarily driven by collapse, we expect the kinetic energy $\mathcal{T}$ should approach the negative of the potential energy, $-\mathcal{W}$. By contrast, a system in virial equilibrium between gravity and kinetic energy obeys $\mathcal{T} = -\mathcal{W}/2$. It is most common to express these relations in terms of the virial ratio $\alpha_\mathrm{vir} \approx 2\mathcal{T}/|\mathcal{W}|$ (we give a more precise definition below), and in this language free-fall collapse corresponds to $\alpha_\mathrm{vir} = 2$ and equilibrium to $\alpha_\mathrm{vir} = 1$. Thus sufficiently precise measurements of the virial ratio \reply{could in principle} tell us whether the motion in a given structure should be attributed to global collapse or not.

In practice, however, \reply{this method does not work. In part this is because the factor of two difference is for idealized collapse scenarios, and depending on the flow geometry the actual value can be significantly closer to unity \citep[e.g.,][]{Vazquez-Semadeni07a, Noriega-Mendoza18a}. More importantly, though,} there are \reply{also} no realistic prospects for measuring virial ratios to \reply{even factor-of-two} precision. In part this is due to observational uncertainties, most prominently the systematic errors associated with converting CO line intensities into estimates of gas surface density \citep{Bolatto13a}. While this uncertainty can be overcome using much more accurate measurements of gas mass from dust  \citep[e.g.,][]{Alves01a, Lombardi14a}, there are also more fundamental problems that are less easily resolved. Calculation of the potential energy to better than a factor of two accuracy requires knowledge of the three-dimensional geometry, which we lack. In extragalactic data we also usually lack knowledge of magnetic fields and their possible contribution to the energy balance. Finally, we do not precisely know the pressures exerted on molecular clouds' surfaces by their atomic envelopes, which also potentially affect the virial ratio. Consequently, it has long been recognized that measurement of the virial ratios of clouds cannot by themselves tell us whether those clouds are in a state of global collapse or not \citep[e.g.,][]{Ballesteros-Paredes06a, Ballesteros-Paredes11a}.

However, recent advances in observational capabilities, combined with the new technique we introduce here, offer the opportunity to overcome this barrier. The central idea behind our technique, which we seek to demonstrate in this paper, is that, while one cannot determine if a structure is globally collapsing based on a single measurement of $\alpha_\mathrm{vir}$ due to the systematic uncertainties we have just discussed, measurements of how $\alpha_\mathrm{vir}$ \textit{changes} with size or density scale within a single object are also diagnostic of the presence or absence of global collapse, and these can be measured with far fewer uncertainties. That is, while \textit{absolute} energies are significantly uncertain, the \textit{relative} energies contained within structures of different length or size scales in a single data set can be measured with much less uncertainty, and allow us to measure with reasonable confidence whether $\alpha_\mathrm{vir}$ is increasing, decreasing, or remaining constant as we select progressively larger or smaller areas or volumes within a GMC. This trend, in turn, is diagnostic of the presence or absence of a global collapse.

Prior to the last five years this point would have been of purely theoretical interest, because the data required to measure $\alpha_\mathrm{vir}$ across a wide range of scales were not available. Within the Galaxy we have been able to achieve the required dynamic range in CO surveys for some time, and \citet{Peretto23a} provide measurements of $^{13}$CO and N$_2$H$^+$ in a sample of Galactic molecular and infrared dark clouds covering a wide range of scales that anticipates much of the work we will present here. However, Galactic measurements suffer from the usual complications of uncertain distances and line-of-sight confusion that make it difficult to accurately or self-consistently measure the basic cloud properties of size, mass, and surface density. Moreover, due to the large extents on the sky of nearby molecular clouds, time constraints have limited acquisition of the combined dust and CO observations needed for accurate simultaneous measurements of mass and kinematics to only a few examples \citep[e.g.,][]{Lewis21a}. For more distant Galactic clouds the problem of the large angular size is less severe, but this is replaced by the problem that it becomes difficult to measure masses using dust due to line-of-sight confusion -- see \citeauthor{Peretto23a} for an extensive discussion.

These factors favor the use of extragalactic observations, which largely eliminate the relative uncertainties in GMC distances and can provide observing sightlines that traverse a much smaller distance through a molecule-rich Galactic midplane than observations of distant clouds within the Milky Way disk, thereby reducing the problem of line-of-sight confusion. However, the depth and spatial resolution attained in most earlier extragalactic CO surveys were such that the dynamic range in both cloud size and column density were too limited to trace out structures over a large dynamic range. Moreover, extragalactic dust data covering wide areas were until recently scarce. This has begun to change thanks to the deployment of wide-band continuum receivers on millimeter- and submillimeter-wave interferometers, which enable resolved measurements of dust emission from individual GMCs in nearby galaxies. Indeed, a recent simultaneous CO and continuum survey of the Andromeda galaxy (M31) with the SMA has achieved the resolution and depths necessary to both resolve dust and CO emission and trace CO emission to the very physical edges of a GMC in that galaxy \citep[][]{Forbrich20a, Viaene21a, Lada24a}. This work illustrates the potential of deep interferometric CO observations to measure the energy budgets across individual GMCs, not just at a single size or column density scale, but over a broad dynamic range in these quantities.

In this paper we first demonstrate the assertion made above that a measurement of how $\alpha_\mathrm{vir}$ changes with scale can be used as a tool for distinguishing clouds that are in local versus global collapse, and we then apply this insight to recent data. We explain the intuitive idea behind our method in \autoref{sec:dva}, and validate it against both analytic solutions and full numerical simulations in \autoref{sec:validation}. We then consider observational uncertainties and complications in \autoref{sec:observations}, and apply the method to a recent data set for the molecular clouds of Andromeda in \autoref{sec:andromeda}; a complete description of this data set, and further applications of the method we present here to it, appears in a companion paper (Lada et al., 2025, submitted). Finally, we summarize and discuss future prospects in \autoref{sec:conclusion}.

\section{Differential virial analysis: motivation and key concepts}
\label{sec:dva}

\subsection{The problem with traditional virial analysis}
\label{ssec:problem_traditional}

Virial analysis involves using observable quantities to estimate the various terms that appear in the virial theorem describing the second derivative of a cloud's moment of inertia. In its Eulerian form, and dropping the magnetic and radiation terms for simplicity, the theorem reads \citep{McKee92a}
\begin{equation}
    \frac{1}{2} \ddot{I} = 2\left(\mathcal{T}-\mathcal{T}_S\right) + \mathcal{W} - \frac{1}{2}\frac{d}{dt} \int_S \left(\rho\mathbf{v}r^2\right) \cdot d\mathbf{S},
    \label{eq:virial}
\end{equation}
where $S$ is the virial surface and $V$ is the volume interior to that surface, and the various terms appearing are
\begin{eqnarray}
    I & = & \int_V \rho r^2 \, dV \\
    \mathcal{T} & = & \int_V \left(\frac{1}{2}\rho v^2 + \frac{3}{2} P\right) \, dV \\
    \mathcal{T}_S & = & \int_S \mathbf{r}\cdot\boldsymbol{\Pi} \cdot d\mathbf{S} \\
    \mathcal{W} & = & -\int_V \rho\mathbf{r}\cdot\nabla\phi\, dV.
\end{eqnarray}
Here $I$ is the moment of inertia of material within the virial surface, $\mathcal{T}$ is the kinetic plus thermal energy of this material, $\mathcal{T}_S$ measures the momentum flux across the surface, and $\mathcal{W}$ is the gravitational binding energy of the material; in turn, $\rho$ is the matter density, $\mathbf{v}$ is the velocity, $P$ is the thermal pressure, $\Pi_{ij}=\rho v_i v_j + P \delta_{ij}$ is the Reynolds stress tensor, and $\phi$ is the gravitational potential.

When making observational estimates of these terms, it is common to make some simplifying assumptions. Specifically, we can choose the virial surface to be a sphere of radius $R$, and then assume that (1) the material inside that surface is isothermal with sound speed $c_s$, (2) the gravitational potential is solely due to matter within that surface, and (3) there is negligible mass flow across the surface. Under these assumptions we can drop the final $d/dt$ term in \autoref{eq:virial} and rewrite the remaining terms as
\begin{eqnarray}
    \mathcal{T} & = & \frac{3}{2} M \sigma^2 \\
    \mathcal{T}_S & = & 4\pi R^3 P_S \\
    \mathcal{W} & = & -a\frac{G M^2}{R},
\end{eqnarray}
where $M$ is the total mass inside the virial surface, $\sigma^2 \equiv \langle v^2\rangle + c_s^2$ is the 1d mass-weighted total thermal ($c_s^2$) plus non-thermal velocity dispersion ($\langle v^2\rangle$, with the angle brackets indicating a mass-weighted average), $P_S$ is the pressure at the virial surface, and $a$ is a constant of order unity that depends on the distribution of matter inside the virial surface; for uniform density $a=3/5$.

For an object in virial balance, $\ddot{I} = 0$, we can then rewrite the virial theorem in the form
\begin{equation}
    \alpha_\mathrm{vir} - \alpha_S = 1,
\end{equation}
where we have defined the virial parameter \citep{Bertoldi92a} and an analogous surface term
\begin{eqnarray}
    \alpha_\mathrm{vir} & = & -\frac{2\mathcal{T}}{\mathcal{W}} 
    \label{eq:avir}
    \\
    \alpha_S & = & -\frac{2\mathcal{T}_S}{\mathcal{W}}.
\end{eqnarray}
Since $\mathcal{W}$ is not an easily-observable quantity due to its dependence on the unknown 3D geometry, it is conventional to define an observable virial parameter
\begin{equation}
    \alpha_\mathrm{vir,obs} = \frac{5\sigma^2/R}{\pi G\Sigma},
    \label{eq:avir_obs}
\end{equation}
where $\Sigma = M/\pi R^2$ is the mean surface density within some radius $R$, and to evaluate $\mathcal{W}$ we have adopted the value of $a$ for a uniform sphere. In this form the virial parameter is in principle expressed purely in terms of observable quantities, and thus can serve as a measurable estimator for the relative importance of self-gravity and surface pressure in determining the dynamics of material inside the virial surface: if $\alpha_\mathrm{vir,obs} = 1$, then surface forces are entirely unimportant, while if $\alpha_\mathrm{vir,obs} \gg 1$ then either surface forces are dominant or the cloud is far from virial equilibrium (or magnetic forces described by terms that we have dropped are dominant). This relationship is commonly-expressed via a plot of cloud properties in the plane of $\sigma^2/R$ versus $\Sigma$ \citep[e.g.,][]{Heyer09a, Field11a}, to which we will refer henceforth as the virial diagram; in this plane, $\alpha_\mathrm{vir,obs} = 1$ appears as line of slope unity, and thus the gravitational binding and importance of surface forces can be read off from whether a given object lies near or far above this line.

However, as already discussed in \autoref{sec:intro}, the difference in $\alpha_\mathrm{vir}$ between a cloud in virial balance confined entirely by self-gravity and one that is in a state of free-fall collapse after starting from rest -- which has $\mathcal{T} = -\mathcal{W}$ and thus $\alpha_\mathrm{vir} = 2$ -- is small; indeed, $\alpha_\mathrm{vir} = 2$ could \textit{also} correspond to a cloud that is marginally bound but also pressure-confined, $\alpha_S = 1$. Moreover, $\alpha_\mathrm{vir,obs}$ it not exactly the same as $\alpha_\mathrm{vir}$, since we have had to assume a value of $a$, our measurements of projected quantities invariably include some material that is along the line of sight but not within our intended virial surface (even neglecting other observational uncertainties), and in addition we have neglected other terms that appear in the full virial theorem, most notably magnetic forces and mass flux across the virial surface. As a result of all these systematic effects, which are certainly at least at the factor of $\approx 2$ level, it is not possible to use measurements of $\alpha_\mathrm{vir,obs}$ to distinguish collapsing from non-collapsing clouds. 

\subsection{The central idea for differential virial analysis}
\label{ssec:central_idea}

While the issues summarized in \autoref{ssec:problem_traditional} are insuperable when it comes to using measurements of the absolute value of the virial parameter to distinguish static from collapsing structures, there is another important difference between such structures: as we consider successively smaller virial surfaces within the same structure, the \textit{relative} sizes of $\alpha_\mathrm{vir}$ and $\alpha_S$ change differently.

As an extreme example, consider the Sun. If we place our virial surface at $R_\odot$, then we will find that $\alpha_\mathrm{vir} = 1$ and $\alpha_S = 0$ to high accuracy, since there is negligible pressure at the Sun's surface, but if we consider a succession of virial surfaces with radii $R \ll R_\odot$, then as $R$ becomes smaller we will find that $\alpha_S$ increases and $\alpha_\mathrm{vir}$ decreases, because surface pressure is increasingly important and self-gravity increasingly unimportant. Indeed, for conditions at the Solar center, $\rho\approx 160$ g cm$^{-2}$, $T\approx 1.6\times 10^7$ K, we have $\alpha_\mathrm{vir} \approx 800 (R/0.01 R_\odot)^{-2}$. As we shall show below more generally, this behavior of $\alpha_\mathrm{vir}$ and $\alpha_S$ both rising as one moves inward are characteristic of objects that are not in a state of global collapse; indeed, exactly this phenomenon is at the heart of models of turbulence-regulated star formation, which all share as a common assumption that molecular clouds are bound at their largest scales, but that any given randomly-selected sub-region within one is likely to be unbound \citep[e.g.,][]{Krumholz05c, Hennebelle11b, Federrath12a, Hopkins13a}.

By contrast, in gas clouds undergoing global collapse, every structure approaches a state in which its kinetic energy is predominantly fed by the gravitational potential energy released during the collapse, and thus on all scales we should expect $\mathcal{T} \sim -\mathcal{W}$ and therefore $\alpha_\mathrm{vir} \sim 2$. This is seen, for example, in simulations by \citet{Camacho20a, Camacho23a}, where in collapsing clouds one finds that nested gas structures defined by different density thresholds have virial ratios that are nearly independent of the density used to define them, at least down to mass scales $\lesssim 1$ M$_\odot$. This is as expected, since in such a structure surface pressure forces are never important. We again demonstrate this proposition more generally below. In a locally-collapsing cloud, by contrast, this state is only reached on the very smallest scales where local collapse is occurring; over most of a cloud's dynamic range we expect the virial ratio to rise as we go to smaller scales. Thus globally-supported (but locally-collapsing) and globally-collapsing structures differ in how $\alpha_\mathrm{vir}$ changes as we zoom in to progressively-smaller scales. In terms of the virial diagram, we expect successively-smaller regions within collapsing structures to remain relatively close to the $\alpha_\mathrm{vir,obs} = 1$ line of slope unity, while supported structures should curve systematically upward, to higher $\sigma^2/R$, at higher surface density $\Sigma$, as we probe systematically less-bound parts of the structure, until eventually we zoom in to such small scales that we identify regions of local collapse and $\alpha_\mathrm{vir,obs}$ goes back down.\footnote{A minor subtlety worth pointing out is that, for this purpose, it is important that we not carry out this analysis exclusively on regions centered around zones of active star formation, since these are biased to represent regions of collapse. We must instead select the analysis regions in a way that is not biased by selecting on tracers of star formation, and instead selects regions of the cloud at equal surface density regardless of whether they are star-forming or not.}

The main advantage of using this feature rather than the absolute location relative to the $\alpha_\mathrm{vir,obs} = 1$ line is the greatly reduced sensitivity to systematic uncertainties. For example, unless there is reason to believe that the internal mass distribution, geometry, or relative importance of magnetic fields change dramatically with scale, such effects should not systematically change the shape traced out in the virial plot for a given structure, even if they do change the location. Thus the shape in the virial diagram is a much more robust diagnostic than the location, and can potentially distinguish between global and local collapse. We discuss systematic uncertainties further in \autoref{sec:observations}.

\section{Validation of the method}
\label{sec:validation}

We now seek to test the intuitive idea developed in the previous section -- that the trajectory that a given structure traces out in the virial diagram is capable of distinguishing collapsing from supported clouds -- against theoretical models. We first do so using analytic spherically-symmetric profiles for which we can perform exact computations in \autoref{ssec:polytropes}, and then in \autoref{ssec:numerical_tests} we use full numerical simulations.

\subsection{Demonstration for analytic profiles}
\label{ssec:polytropes}

\subsubsection{Hydrostatic polytropes}
\label{sssec:hydrostatic}

As our first example we consider hydrostatic polytropic spheres. Polytropes have the advantage that they can be either supported (non-singular) or on the verge of collapse (singular), and thus using them as our example allows us to compare the two cases on equal footing. For the purposes of our numerical computations, we use the formulation of polytropic systems provided by \citet{McKee99b}. Most of the standard analytic profiles commonly-used in studying star formation, for example the Bonnor-Ebert sphere \citep{Ebert55a, Bonnor56a} and the singular isothermal sphere \citep{Shu77a}, are special cases of this general framework. In the \citeauthor{McKee99b} framework, the equation of hydrostatic balance for a spherical distribution of gas obeying the polytropic equation of state $P\propto \rho^{\gamma_p}$ for the gas pressure $P$ and density $\rho$ is reduced to a pair of ordinary differential equations for the homology variables $\mu$ and $\lambda$, describing a dimensionless mass and radius:
\begin{eqnarray}
    \frac{d\ln\mu}{d\chi} & = & 4\pi\gamma_p \frac{\lambda^4}{\mu^2} - \frac{1}{2}\left(4-3\gamma_p\right) 
    \label{eq:mu} \\
    \frac{d\ln\lambda}{d\chi} & = & \gamma_p \frac{\lambda}{\mu} - \frac{1}{2}\left(2-\gamma_p\right),
    \label{eq:lambda}
\end{eqnarray}
where the independent variable $\chi = \ln(\rho_0/\rho)$ is the dimensionless density contrast relative to some reference density $\rho_0$.

Non-singular solutions to these equations obey the inner boundary condition $\lim_{\chi\to \chi_\mathrm{c}} (\mu/\lambda^3) = 4\pi/3$, where the subscript c indicates quantities evaluated at the cloud center. One may integrate the equations to arbitrary $\chi$, but the structure is unstable once $\mu^2/\lambda^4 > 8\pi \gamma_p/(4-3\gamma_p)$. The singular case corresponds to solutions where $\lambda$ and $\mu$ take on constant values that depend only on $\gamma_p$ (see Equations 24 and 25 of \citeauthor{McKee99b}). In either the singular or non-singular cases, one can then convert back from homology to physical variables using the relations
\begin{eqnarray}
    \rho & = & \rho_0 e^{-\chi} \\
    \sigma & = & \sigma_0 e^{-(\gamma_p-1)\chi/2} \\
    r & = & \lambda \frac{\sigma}{\sqrt{G\rho}} \\
    M & = & \mu \frac{\sigma^3}{\sqrt{G^3 \rho}},
\end{eqnarray}
where $\sigma = \sqrt{P/\rho}$, $\sigma_0$ is the reference velocity dispersion defined at the same location as $\rho_0$, and $M$ is the mass enclosed within some radius $r$.

Given a choice of $\gamma_p$ and a corresponding set of solutions $\mu(\chi)$ and $\lambda(\chi)$ -- either singular solutions where $\mu$ and $\lambda$ are constants or non-singular ones where they are functions of $\chi$ derived from a numerical integration of \autoref{eq:mu} and \autoref{eq:lambda} -- we can compute the true and observer's virial parameters as a function of projected column density. Specifically, we define the mean column density and mass-weighted mean square velocity dispersion within some projected radius $R$ as
\begin{eqnarray}
    \Sigma & = & \frac{1}{\pi R^2} \int_0^{r_s} 4\pi r^2 \rho f_A \, dr 
    \label{eq:Sigma} \\
    \left\langle\sigma^2\right\rangle & = & \frac{1}{\pi R^2 \Sigma} \int_0^{r_s} 4\pi r^2 \rho \sigma^2 f_A \, dr,
    \label{eq:sigma2}
\end{eqnarray}
where in the expression above $r_s$ is the radius of the cloud surface, beyond which the density is assumed to be negligible, and $f_A = 1 - \sqrt{1 - \min(R^2/r^2,1)}$ is the fraction of the surface area of a sphere of radius $r$ that lies at cylindrical radius $<R$. For comparison we also compute the true virial parameter within a spherical radius equal to $R$ from the true kinetic and potential energies
\begin{eqnarray}
    \mathcal{T} & = & 4\pi \int_0^R r^2 \rho \left(\frac{3}{2}\sigma^2\right) \, dr 
    \label{eq:T}
    \\
    \mathcal{W} & = & -4\pi \int_0^R r^2 \rho \left(\frac{G M}{r}\right) \, dr.
    \label{eq:W}
\end{eqnarray}

\begin{figure}
    \includegraphics[width=\columnwidth]{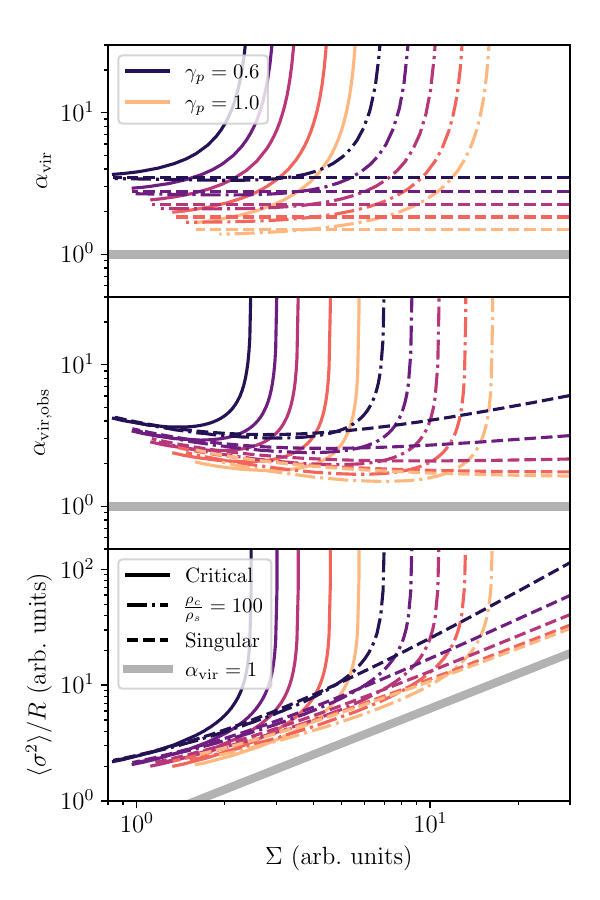}
    \caption{Results for hydrostatic polytropes. Panels show, from top to bottom, the true virial parameter $\alpha_\mathrm{vir}$ (\autoref{eq:avir}), observer's virial parameter $\alpha_\mathrm{vir,obs}$ (\autoref{eq:avir_obs}), and vertical axis of the virial diagram ($\langle\sigma^2\rangle /R$, \autoref{eq:sigma2}) as a function of projected column density $\Sigma$ (\autoref{eq:Sigma}). Lines from dark to light show $\gamma_p = 0.6$ to $1.0$ in steps of 0.1; solid lines show critically stable non-singular polytropes, dotted lines show supercritical non-singular polytropes with center-to-edge density ratios of 100, and dashed lines show singular polytropes; all solutions are expressed in a dimensionless unit system where $G=1$ and the density and velocity dispersion at the cloud edge are also unity. The thick gray line corresponds to $\alpha_\mathrm{vir} = 1$ in all panels.
    \label{fig:avir_polytrope}
    }
\end{figure}

We carry out this calculation for a range of $\gamma_p$ values from $\alpha_p = 0.6 - 1.0$ and plot the resulting true and observable virial parameters, and the virial diagram, as a function of projected column density in \autoref{fig:avir_polytrope}.\footnote{Our maximum value of $\gamma_p = 1.0$ corresponds to the classical Bonnor-Ebert sphere for the non-singular case, and the static singular isothermal sphere for the singular case. We do not consider values of $\gamma_p$ larger than unity because such solutions have velocity dispersions that decrease rather than increase with length scale, contrary to the observed linewidth-size relation. We choose a minimum $\gamma_p = 0.6$ because this gives $\alpha_\mathrm{vir} \approx 3$, and $\alpha_\mathrm{vir}$ rises sharply for smaller values of $\gamma_p$ because such polytropes have pressures that rise strongly with radius in a way that makes them essentially pressure-confined structures, with increasingly small contributions from gravity.} We plot three cases: critical non-singular polytropic spheres that we integrate from their centers to the density contrasts at which they become unstable (solid lines), supercritical non-singular spheres with a center-to-edge density ratio of 100, and singular spheres that are truncated at an arbitrary radius. In all cases we define the unit system for the calculation implicitly by setting $\rho_0 = 1$ and $\sigma_0 = 1$ at the cloud surface, and setting $G=1$; the leftmost points in the curves shown correspond to placing the virial surface at the cloud physical surface.

The results shown in \autoref{fig:avir_polytrope} are consistent with the intuitive expectation developed in \autoref{ssec:central_idea}: the singular cases where the density rises without limit have constant true virial parameters, and nearly constant observed virial parameters; the small deviations between true and observed virial parameter are a result of projection effects combined with our having broken the self-similarity of the results by truncating the profile at a finite radius. By contrast, the non-singular cases have virial profiles that curve upward at high column densities. Thus the shape of the virial parameter versus surface density curve strongly distinguishes between structures that are self-gravitating only at their full size but not in small portions of their interiors and structures that are gravity-confined at all radii.

\subsubsection{Collapsing polytropes}

The hydrostatic polytropic solutions we have just considered demonstrate that differential virial analysis can distinguish internally-supported from singular hydrostatic structures. We now demonstrate that this trends continues to dynamically collapsing structures. For the singular case, we use the analytic solutions for the collapse of singular polytropes provided by \citet{McLaughlin97a}, which provide density $\rho$, velocity dispersion $\sigma$, and radial velocity $v_r$ as a function of radius. For the non-singular cases exact analytic solutions are not known for general $\gamma_p$, but we rely on the fact that at late times the solutions approach the free-fall profiles derived by \citet{Hunter62a}, \citet{Larson69a}, and \citet{Penston69b, Penston69a}. The fundamental result for these solutions, which we take from these papers, is that a shell of material with starting radius $r_0$ at $t=0$ will at some later time $t$ have fallen to a radius $r$ given implicitly by
\begin{equation}
    \xi + \frac{1}{2}\sin 2\xi = \frac{\pi}{2} \tau
    \label{eq:hunter}
\end{equation}
where the dimensionless variables $\xi$ and $\tau$ are defined implicitly by $r/r_0 = \cos^2\xi$ and $\tau = t/t_\mathrm{ff}$ where $t_\mathrm{ff} = (\pi/2) \sqrt{r_0^3/2 G M}$ and $M$ is the mass interior to the shell of mass with initial radius $r_0$. The radial velocity of this shell of material is
\begin{equation}
    v_r = -\sqrt{2 G M\left(\frac{1}{r}-\frac{1}{r_0}\right)}.
\end{equation}
Given an initial mass profile $M(r_0)$, we can use these relations to find the mass profile $M(r,t) = M(r_0(r,t))$ at any later time $t$, where $r_0(r,t)$ is the value of $r_0$ that solves \autoref{eq:hunter} for the specified $(r,t)$. The density profile at this time is then given by $\rho(r,t) = (1/4\pi r^2) (\partial M/\partial r)$, and for polytropic material the velocity dispersion is then $\sigma = \sigma_0 (\rho/\rho_0)^{(\gamma_p-1)/2}$.

Given these solutions for the density, velocity dispersion, and radial velocity as a function of radius in collapsing polytropes, we can compute the projected column density $\Sigma$ and the true gravitational potential energy $\mathcal{W}$ exactly as for the hydrostatic case (\autoref{eq:Sigma} and \autoref{eq:W}). For the true kinetic energy and projected velocity dispersion, we must include the contribution from bulk radial motion, and therefore we generalize these definitions to
\begin{eqnarray}
    \mathcal{T} & = & 4\pi \int_0^R r^2 \rho \left(\frac{3}{2}\sigma^2 + \frac{1}{2} v_r^2\right)\, dr \\
    \left\langle\sigma^2\right\rangle & = & \frac{1}{\pi R^2 \Sigma} \int_0^{r_s} \int_0^{\theta_\mathrm{max}} 4\pi r^2 \rho 
    \nonumber \\
    & & \qquad 
    \left[\sigma^2 + v_r^2\cos^2\theta\right] \sin\theta\, d\theta \, dr, 
\end{eqnarray}
where $\theta_\mathrm{max} = \sin^{-1} [\min(R/r, 1)]$. Note that from 0 to $\theta_\mathrm{max}$ is the range of angles that such that material at position $(r,\theta)$ lies within cylindrical radius $R$ of the central axis, and that the factor $\cos^2\theta$ here ensures that we only count the line-of-sight component of $v_r$ in our estimate of $\langle\sigma^2\rangle$.

\begin{figure}
    \includegraphics[width=\columnwidth]{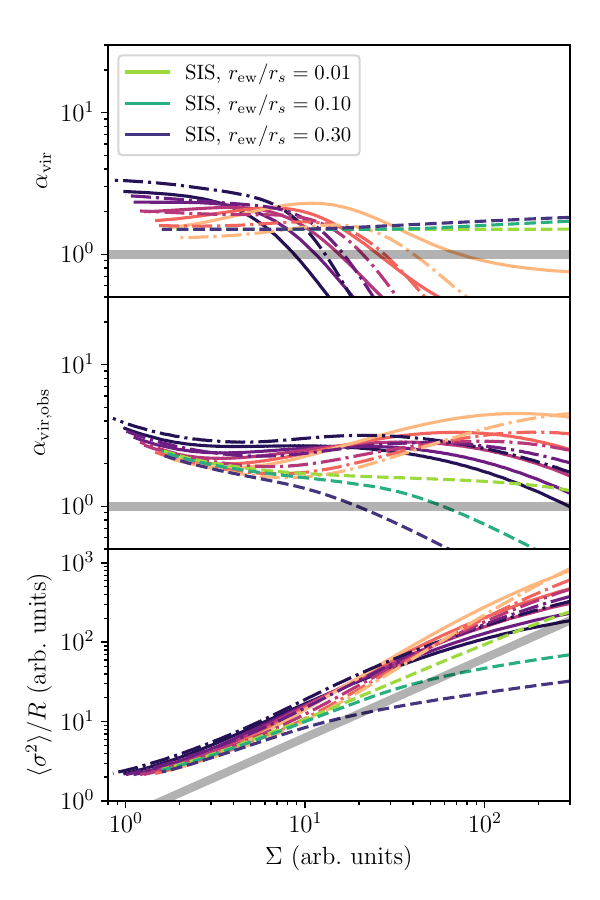}
    \caption{Same as \autoref{fig:avir_polytrope}, but now for collapsing rather than hydrostatic polytropes. Lines colored from orange to black are for the same cases as in \autoref{fig:avir_polytrope}: non-singular polytropes with a range of polytropic indices $\gamma_p$ that are initially either critical (solid lines) or supercritical with density contrasts of 100 (dot-dashed lines). In all cases these are shown just after the instant of singularity formation, at $t=1.01 t_\mathrm{ff,c}$, where $t_\mathrm{ff,c}$ is the central free-fall time prior to collapse. The green-to-purple lines show a singular isothermal sphere, $\gamma_p = 1$, at three phases during its collapse characterized by $r_\mathrm{ew}/r_s = 0.01$, 0.1, and 0.3, where $r_\mathrm{ew}$ and $r_s$ are the radii of the expansion wave and the cloud surface, respectively.
    \label{fig:avir_polytrope_collapse}
    }
\end{figure}

We show $\alpha_\mathrm{vir}$, $\alpha_\mathrm{vir,obs}$, and the virial diagram as a function of projected column density $\Sigma$ for these collapse cases in \autoref{fig:avir_polytrope_collapse}. For the non-singular cases, we show the state at a time just after singularity formation: $t = 1.01t_\mathrm{ff,c}$, where $t_\mathrm{ff,c}$ is the central free-fall time of the initial, hydrostatic state. For the singular case, we show three different times, where we measure time by the ratio of the radius of the expansion wave $r_\mathrm{ew}$ to the radius the cloud surface $r_s$. We show only the singular isothermal sphere case $\gamma_p = 1$ to minimize clutter, but other values of $\gamma_p$ are not qualitatively different.

The primary conclusion to be drawn from \autoref{fig:avir_polytrope_collapse} is that they are consistent with our expectations regarding the behavior of collapse solutions in the virial diagram. That is, if we focus on the virial diagram (third panel), we see that collapsing solutions are either parallel to the $\alpha_\mathrm{vir,obs} = 1$ line or curve downward towards it. None of them show the upward curve that is apparent for the hydrostatic, non-singular solutions shown in \autoref{fig:avir_polytrope}. These plots therefore support the intuitive argument presented in \autoref{ssec:central_idea}: that we can distinguish supported, non-collapsing configurations from non-supported or collapsing ones based on the fact that, in the virial diagram, the former curve upward away from the virial line as we move to higher $\Sigma$, while the latter curve downward.

\subsection{Tests on numerical simulations}
\label{ssec:numerical_tests}

We next test our intuitive idea on realistic 3D simulations. For this purpose, we draw an example from the Catalog of Astrophysical Turbulence Simulations (CATS; \citealt{Burkhart20a}): the simulations of \citet[analysis for which appears in \citealt{Burkhart15b}]{Collins12a}. These simulations in particular are useful for our purposes because, as \citeauthor{Collins12a} show, they pass through both supported and collapsing phases. The simulations are of the driven-box type: they begin with a periodic box in which self-gravity is turned off and turbulence is driven until it reaches statistical steady-state. In this case, the driving rate is set so that, once the simulations settle to steady-state (and before gravity is turned on), the virial parameter is unity and the sonic Mach number is nine. Once the simulations have run for several crossing times without gravity, driving ceases, gravity is turned on, and the simulation is allowed to proceed. At early times, $\sim 0.1 t_\mathrm{ff}$, where $t_\mathrm{ff}$ is the mean-density free-fall time, the cloud is turbulence-supported, but as the simulation time approaches $\sim t_\mathrm{ff}$ it becomes increasingly dominated by collapse; \citeauthor{Collins12a} show that this transition is reflected in numerous statistical properties of the density, velocity, and magnetic field structure. These simulations therefore offer us the opportunity to test differential virial analysis within a single, self-consistent framework that displays both of the types of behavior in which we are interested. For the remainder of our analysis, we therefore use the set of simulations from \citeauthor{Collins12a} with plasma $\beta=0.2$, for which CATS provides snapshots of density and velocity at $256^3$ resolution (reduced from the native adaptive mesh resolution) at $t/t_\mathrm{ff} = 0$, 0.1, 0.3, and 0.6.

To test differential virial analysis on this data set, we process each snapshot as follows. First, we compute the projected column density and mass-weighted mean velocity dispersion in the $x$ direction as
\begin{eqnarray}
    \label{eq:Sigmapix}
    \Sigma & = & \int_0^1 \rho \, dx \\
    \left\langle v_x\right\rangle & = & \frac{1}{\Sigma} \int_0^1 \rho v_x \, dx \\
    \sigma^2 & = & \frac{1}{\Sigma} \int_0^1 \rho \left[1 + \left(v_x - \left\langle v_x\right\rangle\right)^2\right] \, dx,
    \label{eq:sigma2pix}
\end{eqnarray}
where $\rho$ and $v_x$ as the gas density and $x$-velocity, and all quantities are expressed in code units whereby the simulation domain extends over the range $(0,1)$ in all cardinal directions and the gas sound speed and mean density are both unity.\footnote{In these code units, $G = (5/3)\times 9^2$.} We also compute the analogous projections in the $y$- and $z$-directions.

\begin{figure}
    \includegraphics[width=\columnwidth]{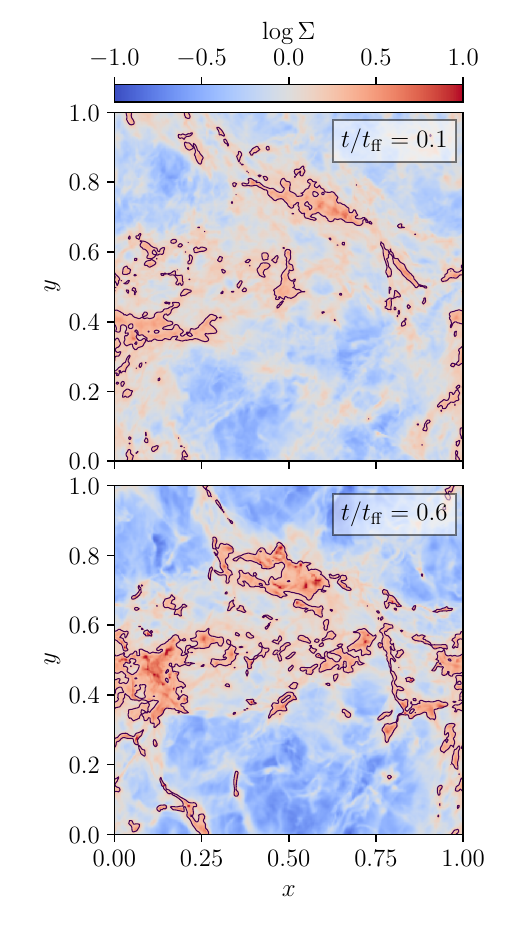}
    \caption{Example of contour analysis on the \citet{Collins12a} simulations. The background images in the upper and lower panels show the column density $\Sigma$ projected in the $z$ direction for two snapshots, at $t/t_\mathrm{ff} = 0.1$ and $0.6$, respectively. All quantities are simulation units where the simulation domain size is unity in each direction, and the mean column density through the domain is also unity. The black line shows contours generated at $\log\Sigma = 0.25$; for our analysis we compute the observers virial ratio $\alpha_\mathrm{vir,obs}$ inside each contour that contains $\geq 30$ simulation pixels.\\
    \label{fig:sim_contour}}
\end{figure}

Next, we define a series of surface density contour levels $\Sigma$, starting from just above the mean surface density of the simulation $\langle\Sigma\rangle = 1$ and increasing to highest surface densities we can resolve, and use flood filling (as implemented in the \textsc{SciKit-Image} package; \citealt{Walt14a}) to identify connected regions of pixels where $\Sigma$ exceeds each contour level in each projection.\footnote{This step requires some care to handle the periodicity of the domain, since there are connected regions that cross the periodic boundary of our projection. Our method for doing so as is follows. First, we tile the projections in a $3\times 3$ grid, and find connected regions on this tiled image. In order to avoid double-counting, we then discard regions whose centroids lie outside the central $1\times 1$ section of the grid representing ``real'' pixels. We also discard any regions whose centroids pass this test but that extend all the way to the edge of the tiled box, since such regions must be larger than fit in our periodic domain and thus we cannot compute meaningful statistics for them.} \autoref{fig:sim_contour} shows an example of the contours we generate via this process for one projection and contour level in two different snapshots. We discard regions containing fewer than 30 pixels (corresponding roughly to a circle of radius 3 pixels) as unresolved, and then for all remaining regions $\mathbf{R}$ we compute a mean column density, squared velocity dispersion, effective radius, and observer's virial parameter as
\begin{eqnarray}
    \left\langle\Sigma\right\rangle & = & \frac{1}{N_\mathrm{pix}} \sum_{i\in \mathbf{R}} \Sigma_i \\
    \left\langle\sigma^2\right\rangle & = & \frac{1}{N_\mathrm{pix} \left\langle\Sigma\right\rangle}\sum_{i\in \mathbf{R}} \Sigma_i \sigma^2_i \\
    R_\mathrm{eff} & = &\sqrt{ \frac{N_\mathrm{pix} A_\mathrm{pix}}{\pi}} \\
    \alpha_\mathrm{vir,obs} & = & \frac{5\left\langle\sigma^2\right\rangle/R_\mathrm{eff}}{\pi G\left\langle\Sigma\right\rangle},
\end{eqnarray}
where $N_\mathrm{pix}$ is the number of pixels in region $\mathbf{R}$, $\Sigma_i$ and $\sigma^2_i$ are the column density and squared velocity dispersion of pixel $i$, and $A_\mathrm{pix} = 1/256^2$ is the area of a pixel.

\begin{figure}
    \centerline{\includegraphics[width=\columnwidth]{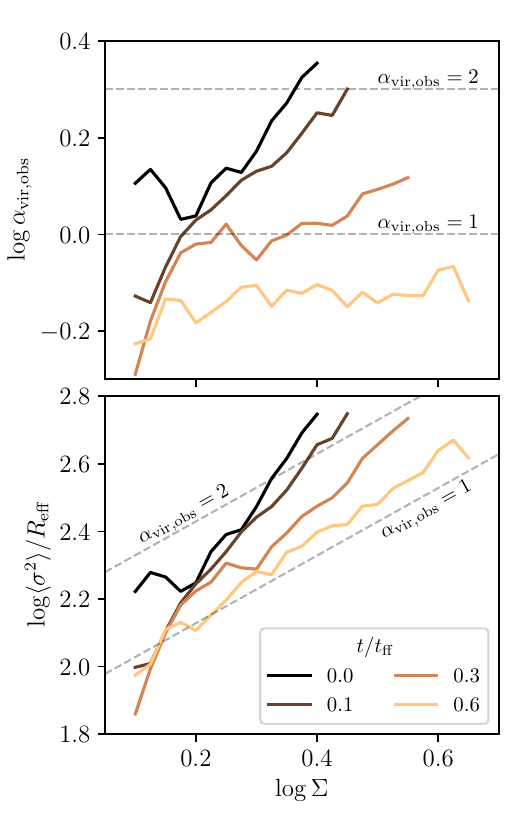}}
    \caption{Differential virial analysis for the simulations of \citet{Collins12a}, which are supported at early times but transition to globally collapsing at later times. Lines show simulation times $t/t_\mathrm{ff} = 0$, 0.1, 0.3, and 0.6, as indicated, progressing from supported to collapsing. The top panel shows the observer's virial parameter $\alpha_\mathrm{vir}$, while the bottom panel shows the virial diagram $\sigma^2/R$ versus $\Sigma$. All quantities are plotted in code units, where the simulation domain size, mean column density, and sound speed are all unity, and all quantities are mass-weighted means over all contours. Dashed gray lines indicate virial parameters $\alpha_\mathrm{vir,obs} = 1$ and 2 to guide the eye. Note that, as the simulation proceeds from supported to collapsing, the virial diagram changes from upward sloping (relative constant $\alpha_\mathrm{vir}$) to flat, the key signature for differential virial analysis.\\
    \label{fig:avir_sim}}
\end{figure}

The outcome of this process is that, for each contour level, we have a set of regions for which we have computed all of the quantities that enter into the observer's virial parameter. For each contour level for which we have at least five such regions, we then compute the mean weighted by region mass, $M = \langle\Sigma\rangle N_\mathrm{pix} A_\mathrm{pix}$. We plot the results as a function of contour level in \autoref{fig:avir_sim}. The key message of this Figure is that the signature on which differential virial analysis relies -- a change from upward-sloping to flat or downward-sloping in the virial diagram as we go from supported to collapsing clouds -- is clearly evident in these simulations. We see that at the earlier times when the cloud is supported, $t/t_\mathrm{ff} = 0$ and $0.1$, the path traced out by structures as we move from low to high surface density threshold $\Sigma$ goes from lower to higher $\alpha_\mathrm{vir,obs}$, yielding a line in the virial diagram (right column) that tilts upward relative to the constant $\alpha_\mathrm{vir}$ line of slope unity. This is exactly the same signature we identified for supported polytropic spheres in \autoref{sssec:hydrostatic}, albeit with significantly less dynamic range than we see for purely analytic solutions.\footnote{We do not see the final return to low virial at the absolute smallest scales around individual collapsing cores simply because such structures are too small to pass our size threshold.} As the state of the cloud transitions from supported to collapsing over time, starting at $t/t_\mathrm{ff} = 0.3$ and very clearly by $t/t_\mathrm{ff} = 0.6$, this relationship flattens and we find paths of slope unity in the virial diagram. Our numerical analysis therefore supports the idea that we can use the shape or slope in the virial diagram to distinguish supported from collapsing structures.

\reply{However, we do caution that we have tested only a single particular set of simulations. It would be useful to repeat the analysis we have provided here on a broader set of simulations to ensure that our conclusions are generalizable. To facilitate such efforts we have made our analysis software freely available (see Software Availability Statement).}

\section{Systematic uncertainties in observations}
\label{sec:observations}

Now that we have validated the concept of differential virial analysis on both analytic profiles and simulations, it is important to ask whether there are likely to be systematics present in real observations that will thwart application of the method to them. We note that, since the simulations to which we have compared in \autoref{ssec:numerical_tests} already include realistic non-spherical 3D cloud shapes and strong magnetic fields, we need not consider those systematics further; our technique survives them. We therefore focus primarily on effects that we have not yet considered in our simulated observations. The observers' virial parameter $\alpha_\mathrm{vir,obs}$ depends on the surface density $\Sigma$ and the velocity dispersion $\sigma$, and so we focus on biases that could cause our estimates of these to be off differentially as a function of size or column density scale within a single cloud, since such effects could mimic or mask the signatures that are apparent in \autoref{fig:avir_sim}. We first consider these effects in general terms, and then perform an explicit check using simulated line observations of our simulations.

\subsection{Surface density bias}

If we estimate the surface density $\Sigma$ from line surveys, we must use a conversion factor such as $\alpha_\mathrm{CO}$, or analogous quantities for other lines, to convert the observable line surface brightness into an estimate of the mass surface density. The usual assumption is that $\alpha_\mathrm{CO}$ is constant within a single GMC, and thus systematic variations in this quantity are a potential source of bias. Of course one obvious way to avoid such a bias is to estimate the mass using an alternative, less-biased tracer such as dust, and that is the approach we will follow in the sample application we present in \autoref{sec:andromeda}. It is, however, useful to ask about the extent to which we expect such a bias to be a concern when such data are not available. There have been a large number of studies of variations in $\alpha_\mathrm{CO}$ on which we can draw to reach some general conclusions on this question \citep[e.g.,][]{Narayanan11a, Narayanan12a, Bolatto13a, Gong20a, Lada20a, Hu22b, Teng23a}. 

By far the most important factor driving $\alpha_\mathrm{CO}$ variations is the metallicity, but this is likely very close to uniform within a single molecular cloud due to efficient mixing during cloud formation \citep{Feng14a, Armillotta18a}, and thus will not introduce a bias in differential virial analysis.  There can, however, be secondary variation of $\alpha_\mathrm{CO}$ with other physical properties that might vary on smaller scales, and two in particular seem possibly important here. One is a chemical effect: at low column densities, not all of the carbon will be locked in CO, reducing the CO luminosity per unit mass. As a result, using CO as a tracer can lead one to underestimate the column density. In the virial diagram (e.g., right column of \autoref{fig:avir_sim}), if we were to ignore this effect and assume fixed $\alpha_\mathrm{CO}$, the effect would be to pull data points at low $\Sigma$ to even lower $\Sigma$, and thus to flatten the observed relation relative to the true one; this could therefore cause us to mistake supported clouds for collapsing ones. However, \citet{Hu22a} find that the effect is only important at column densities below the threshold where dust extinction blocks interstellar far ultraviolet field, which happens at visual extinction $A_V\sim 3$ mag (corresponding to a column density $\Sigma \approx 50$ M$_\odot$ pc$^{-2}$ at Solar metallicity). At higher extinctions $\alpha_\mathrm{CO}$ becomes relatively constant, and thus chemical bias appears unlikely to be important as long as we limit our studies to higher-column regions.

The other potential source of local variation in $\alpha_\mathrm{CO}$ is due to the optical depth of the CO line \citep{Teng23a}, which depends on both the matter column density and the gas velocity dispersion \citep{Narayanan11a, Narayanan12a}. Sightlines with lower optical depth have smaller $\alpha_\mathrm{CO}$, and thus assuming a constant $\alpha_\mathrm{CO}$ would lead to an overestimate of the gas column density on low-optical depth sightlines. The effect of this on the virial diagram depends on whether low optical depths occur predominantly at lower or high surface density $\Sigma$, which in turn depends on the covariance of the velocity dispersion $\sigma$ with $\Sigma$. When comparing GMCs across the face of a galaxy, higher surface densities tend to be correlated with higher velocity dispersions, and while the two effects push in opposite directions -- higher column density increases optical depth, while higher velocity dispersion decreases it -- the latter effect seems predominate. Consequently, \citet{Teng23a} find that GMCs nearer to galactic centres, which have higher surface density, also tend to have lower optical depth and thus lower $\alpha_\mathrm{CO}$.

However, we should not assume that the same trend will apply within a single cloud, which is what matters for our purposes. Indeed, in the data set we examine in \autoref{sec:andromeda}, velocity dispersions are mostly relatively flat or weakly decreasing with column density, suggesting that optical depths and $\alpha_\mathrm{CO}$ will both increase toward the denser parts of clouds. If this is the case, then the simplest assumption of constant $\alpha_\mathrm{CO}$ would lead us to underestimate column densities at high $\Sigma$, steepening the virial diagram and pushing collapsing clouds in the direction of appearing supported. However, we caution that this is a preliminary speculation, since it is based on an extrapolation of observational results derived at galactic scales to far smaller scales. The safest option in any event is to have methods of estimating column density that are more sophisticated than simply assuming constant $\alpha_\mathrm{CO}$.

\subsection{Kinematic bias}
\label{ssec:kinematic_bias}

The other quantity that enters the virial diagram that might be subject to observational biases is the velocity dispersion. It is well known that observations with different molecular tracers often yield somewhat different linewidths along the same line of sight \citep[e.g.,][]{Andre07a, Kirk07a, Muench07a, Rosolowsky08c}. A number of effects potentially contribute to these differences: sub-thermal excitation and thus suppressed emission in low-density gas \citep[e.g.,][]{Offner08a}, opacity broadening due to higher optical depths at line center than on line wings (e.g., \citealt{Hacar16a}), and sensitivity bias whereby noise preferentially masks the weak wings of lines, making them appear narrower. If these effects were to vary systematically from cloud centers to edges, the resulting bias could skew the virial diagram and affect differential virial analysis.

The most comprehensive study of these effects published to date is that of \citet{Yuan20a}, who generate synthetic line position-position-velocity cubes from simulations including realistic treatments of non-LTE excitation, optical depth, and finite signal-to-noise ratio (SNR). They find that low-density tracer lines that are typically measured at high SNR, such as CO $J=1\to 0$, tend to overestimate velocity dispersions, but that the effect is relatively small, only $\approx 20\%$. The bias is greatly reduced for lines that can be excited at moderate density but are less optically thick, for example CO isotopologues. By contrast, dense gas tracers (e.g., N$_2$H$^+$ $J=1\to 0$ and NH$_3(1,1)$, can underestimate velocity dispersions by a similar amount, particularly if observed with low SNR. Based on this analysis, we can note first that these effects are too small to mimic the variation seen in \autoref{fig:sim_contour} -- even if we assume a systematic 20\% error in estimating $\sigma$ at cloud centers versus edges this amounts to less than a factor 1.5 change in $\alpha_\mathrm{vir,obs}$, and the real error is likely to be even smaller because $\approx 20\%$ is the maximum possible error, and it seems unlikely that our systematics will go all the way from zero to the maximum within a single cloud. And finally, as with surface density bias, we can minimize these concerns by measuring velocity dispersions using one of the less-biased tracers such as rarer CO isotopologues. We will again pursue this strategy for our observed data set in \autoref{sec:andromeda}.

\subsection{Tests using simulated observations}

To follow up on the discussion in the previous two sections, we explicitly investigate the effects of bias by performing simulated observations of the same simulations used in \autoref{ssec:numerical_tests} in the $^{12}$CO and $^{13}$CO $J=1\to 0$ lines, the same as those used in the data to which we will apply our model in \autoref{sec:andromeda}. Our pipeline for simulated observations follows the procedure outlined in \citet{Yuan20a} (and earlier in \citealt{Onus18a}) relatively closely, and so we simply summarize it here and refer readers to that paper for full details.

We first use \texttt{DESPOTIC} \citep{Krumholz14b} to generate a table of $^{12}$CO and $^{13}$CO $J=1\to 0$ luminosity per H nucleon as a function of gas density $n_\mathrm{H}$ and velocity gradient $dv/dr$ using the large velocity gradient (LVG) approximation.\footnote{\texttt{DESPOTIC} uses atomic data taken from the Leiden Atomic and Molecular Database \citep[\url{https://home.strw.leidenuniv.nl/~moldata/}]{Schoier05a}; the specific $^{12}$CO and $^{13}$CO data that we use are originally from \citet{Yang10a}.} For each value of $n_\mathrm{H}$ and $dv/dr$ in our table, we solve for the statistical equilibrium (but not thermodynamic equilibrium) level population distribution of both molecules in the LVG approximation, and from these level populations and the corresponding escape probabilities we then compute the luminosity. For the purposes of this calculation we assume that the background material consists of H$_2$ with an ortho-to-para ratio of 0.25 and He at an abundance of 1 He nucleon per 10 H nuclei, at temperature $T_g = 10$ K. The emitting molecules have fixed abundances H nucleon $x_\mathrm{^{12}CO} = 1.6\times 10^{-4}$ and $x_\mathrm{^{13}CO} = x_\mathrm{^{12}CO}/200$ for $^{12}$CO and $^{13}$CO, respectively. Our table covers number densities $n_\mathrm{H}$ from $10^{-2}$ to $10^8$ cm$^{-3}$ and velocity gradients $dv/dr = 10^-3$ to $10^3$ km s$^{-1}$ pc$^{-1}$, with points logarithmically spaced at steps of 0.1 dex in both quantities.

Once we have generated our tables, our next step is to assign a luminosity in both lines to each simulation cell using bicubic interpolation in the log of H number density and velocity gradient; we take the velocity gradient to be equal to the divergence of the velocity field $\nabla \cdot \mathbf{v}$, which we compute for our gridded simulation data using centered differences. Carrying out this step requires assigning dimensional densities and velocities to the dimensionless simulations; we do so by assigning a gas temperature $T_g = 10$ K (corresponding to a sound speed $c_s = 0.19$ km s$^{-1}$) and a mean density $n_\mathrm{H} = 100$ cm$^{-3}$; all remaining dimensional quantities can then be derived from these two (equations 4 - 8 of \citealt{Collins12a}), and in particular this choice makes the simulation box size $13.7$ pc. We find that choosing $n_\mathrm{H} = 10^3$ cm$^{-3}$ instead, the value recommended by \citeauthor{Collins12a}, yields almost identical results. Once we have assigned each cell a luminosity, we next evaluate the column density and velocity dispersion in each pixel using the CO luminosities rather than the true masses, i.e., we replace the expressions given by \autoref{eq:Sigmapix} - \autoref{eq:sigma2pix} with
\begin{eqnarray}
    \Sigma_\mathcal{L} & = & \mathcal{N} \int_0^1 \mathcal{L} \, dx \\
    \langle v_x\rangle_\mathcal{L} & = & \frac{\mathcal{N} \int_0^1 \mathcal{L} v_x \, dx}{\Sigma_\mathcal{L}} \\
    \sigma^2_\mathcal{L} & = & \frac{\mathcal{N} \int_0^1 \mathcal{L} \left[1 - \left(v_x - \langle v_x\rangle_\mathcal{L}\right)^2\right]\, dx}{\Sigma_\mathcal{L}},
\end{eqnarray}
where $\mathcal{L}$ is the luminosity per unit volume in the $^{12}$CO or $^{13}$CO $J=1\to 0$ line determined from our interpolation and $\mathcal{N}$ is a normalization factor. If we were working with dimensional quantities then $\mathcal{N}$ would be set to $\alpha_\mathrm{CO}$, but we instead set $\mathcal{N}$ to enforce that the mean column density $\Sigma_\mathcal{L}$ inferred from CO lines is unity, identical to the true mean column density; this is convenient because it puts the true and CO-inferred column densities in the same unit system, and it is acceptable for our purposes precisely because we do not care about absolute positions in the virial diagram, only about shapes, and thus the value we choose for $\mathcal{N}$ -- which represents a pure translation in the virial diagram -- drops out of our final analysis.

We then repeat the remainder of the analysis in \autoref{ssec:numerical_tests} to derive virial ratios from the simulated observations. We do this in two ways: a ``mixed'' approach whereby we use the velocity dispersions $\sigma^2_\mathcal{L}$ derived from synthetic observations rather than the true ones, but we continue using the true value of the surface density $\Sigma$, and a ``pure line'' approach whereby we use the line-derived values of both $\Sigma_\mathcal{L}$ and $\sigma^2_\mathcal{L}$. The mixed case corresponds roughly to a procedure where we estimate the mass using dust data that is not biased by molecular excitation and radiative transfer effects and thus only kinematic bias is present, and the molecular line data only for kinematics, while the pure line case corresponds to a procedure that is possible using line data only, and where kinematic and surface density bias both occur. (The observational data we present below in \autoref{sec:andromeda} are intermediate between these two regimes, since the masses are derived from line data, but those line data are in turn calibrated against dust.)

\begin{figure}
    \includegraphics[width=\columnwidth]{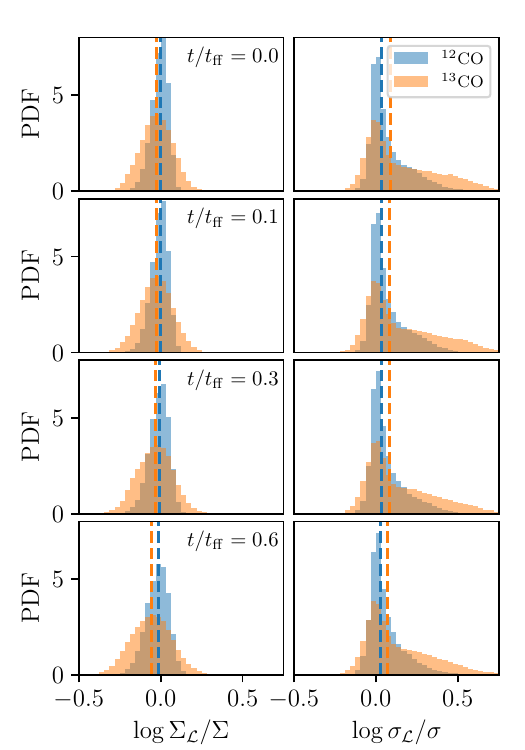}
    \caption{Histogram of ratios $\Sigma / \Sigma_\mathcal{L}$ and $\sigma_\mathcal{L}/\sigma$, where $\Sigma$ and $\sigma$ are the column density and velocity dispersion in projected pixels, quantities subscripted by $\mathcal{L}$ are those derived from synthetic $^{12}$CO $J=1\to 0$ (blue) or $^{13}$CO $J=1\to 0$ (orange), and those with no subscripts are true values. The histograms combine the data projected along the three cardinal directions, and rows show the results at the four times $t/t_\mathrm{ff}$ indicated in each panel. Vertical dashed lines show the median of each distribution.\\
    \label{fig:sigma12_13_ratio}}
\end{figure}

\begin{figure*}
    \includegraphics[width=\textwidth]{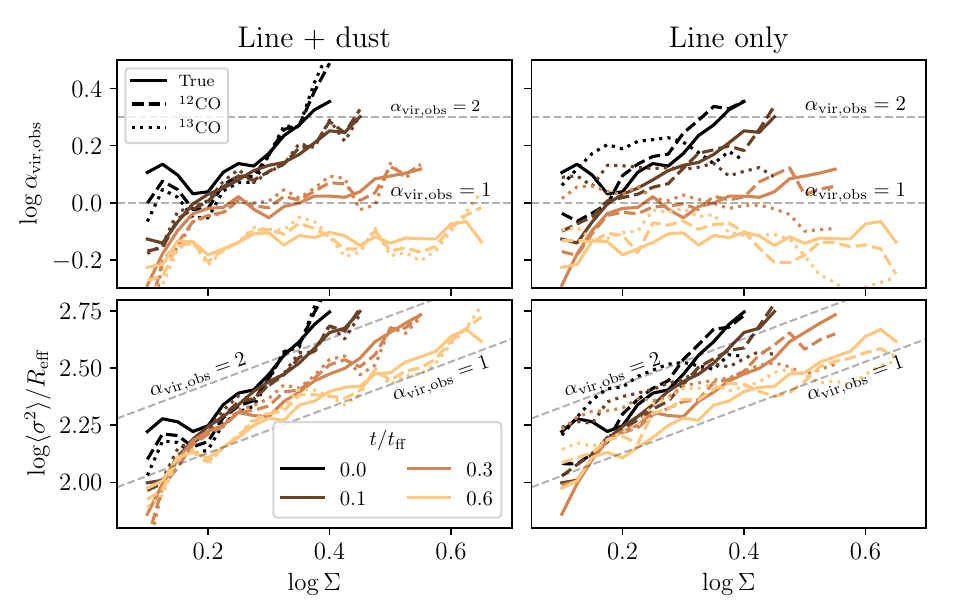}
    \caption{Same as \autoref{fig:avir_sim}, but now comparing results derived from true velocities (solid lines, identical to those shown in \autoref{fig:avir_sim}) to those derived from simulated observations of the $^{12}$CO $J=1\to 0$ (dashed) and $^{12}$CO $J=1\to 0$ (dotted) lines. In the left column we show an analysis using mixed data where we use the line for kinematics but the dust for column densities, while in the right column we show pure line data. In order to facilitate comparison of the curve shapes for true versus $^{12}$CO and $^{13}$CO, we have removed the mean offset between the $^{12}$CO and $^{13}$CO and true lines, so that the means of all the curves shown are identical by construction.\\
    \label{fig:avir_sim_12_13}}
\end{figure*}

\autoref{fig:sigma12_13_ratio} shows the distribution of ratios $\Sigma_\mathcal{L}/\Sigma$ and $\sigma_\mathcal{L}/\sigma$ over all pixels for all cardinal directions and times. For column densities we see that the distribution for $^{12}$CO is relatively narrow at early times, and broadens slightly at later times, with a slight skew to values $<1$; the trends are similar but more exaggerated for $^{13}$CO. Velocities show a more prominent tail skewed to high values, particularly for $^{13}$CO. For all quantities, however, the median error in all quantities is a few $\times 0.01$ dex. We emphasize that this should not be taken as the \textit{absolute} accuracy of CO-derived masses in particular, which is considerably worse; as noted above we have enforced that the mean column density is correct and thus removed the mean error entirely. Instead, this is a measure of the differential error, which is what matters for our proposed analysis method.

To see how these errors propagate through to differential virial analysis, in \autoref{fig:avir_sim_12_13} we reproduce \autoref{fig:avir_sim}, but now using data derived from the simulated observations; the left column shows mixed data where we use lines for kinematics but dust for mass, while the right shows data derived from lines only. We have again removed constant offsets, enabling us to focus on the shapes of the curves only.

We see that for the case of mixed data, the shapes of the differential virial curves are very similar between the true and simulated observation values. The largest difference is at $t=0$, when the simulated observations show a slightly stronger upturn than the true data, but this effect vanishes at all later times. In particular, at $t/t_\mathrm{ff} = 0.3$ and $0.6$, the times when collapse has set in and differential virial analysis detects this by the flattening of the $\alpha_\mathrm{vir}$ versus $\Sigma$ curve, the simulated observations curves show the same flattening as the true data.

For the case of pure line data, we again find that the curves a qualitatively very similar with the exception of $^{13}$CO data at $t/t_\mathrm{ff} = 0$ and 0.1. At these times and for this line, the increase in $\alpha_\mathrm{vir}$ with $\Sigma$ that appears in the true data is lost, and the curve becomes flat. Since this occurs only for the pure line case it must be due to surface density bias, and the cause is easy to understand: $^{13}$CO misses lower-density gas because radiative trapping is weak for this species, and as a result $^{13}$CO molecules are sub-thermally excited in gas with densities below $\sim 10^3$ cm$^{-3}$. Because this bias preferentially affects more diffuse regions, we underestimate the column density more for low-$\Sigma$ regions than for high-$\Sigma$ ones, which in turn boosts $\alpha_\mathrm{vir}$ in low-density gas relative to high-density gas, masking the real rise in $\alpha_\mathrm{vir}$ with $\Sigma$. However, with this understanding, we can see that this bias is only capable of leading to errors in one direction: it can make supported clouds appear collapsing, but not collapsing clouds appear supported -- and we again emphasize that this occurs only for higher-density tracer in the case where we are relying on pure line data, with no dust information. Any time we have dust information, or even $^{12}$CO data, our method is robust, supporting our central contention that differential changes in virial ratio can be measured even when absolute values cannot be.

\section{Application to the giant molecular clouds of Andromeda}
\label{sec:andromeda}

We now provide an initial application of our new technique to the GMCs of Andromeda (M31) using a data set taken from \citet{Lada24a}, which provides deep Submillimeter Array (SMA) measurements of the $^{12}$CO and $^{13}$CO $J=1\to 0$ lines and dust continuum for a sample of $\sim 100$ GMCs; we refer readers to \citeauthor{Lada24a}, to the companion paper to this one \citep{Lada25a}, and to the earlier papers in the series \citep{Forbrich20a, Viaene21a}, for full details on the observational setup and the methods to used to derive velocity dispersions and surface densities. Here we simply note that these methods avoid most of the major systematic uncertainties we identify in \autoref{sec:observations}. In particular, all clouds are at the same distance, eliminating relative distance uncertainties that hinder Galactic observations. The uncertainty due metallicity is minimized due to both the relatively shallow metallicity gradient in M31 and the apparent lack of dependence on metallicity of the CO-to-dust mass conversion factor \citep{Bosomworth25a}. The inclination of M31 is sufficient to minimize confusion from overlapping clouds, and for our analysis we use only GMCs where the $^{12}$CO data are well-fit by a single Gaussian component (which describes the majority of the sample). We measure velocity dispersions from both the $^{13}$CO and $^{12}$CO lines in locations where the former is detected at sufficiently high signal-to-noise ratio (SNR), as described below. Finally, in this data set gas surface densities are self-consistently calibrated from the simultaneous dust continuum observations of a subset of the M31 GMCs themselves \citep{Viaene21a}, meaning that our column densities are derived using a method intermediate between the ``mixed'' and ``line-only'' cases described in the previous section.  We also note that the primary qualitative feature we see in these data -- an upturn in the virial parameter at small radius -- is also seen in the independent measurements of \citet[their Figure 7]{Peretto23a}.

In traditional virial analyses \citep[e.g.,][]{Keto86a, Heyer09a, Faesi18a}, one generally measures the quantities that enter the virial diagram at the lowest detectable emission level. To carry out our differential virial analysis we need instead to construct the virial diagram for the gas within an individual GMC, which we do following the novel methodology described by Lada et al. (2025, submitted). Briefly, for each GMC in the sample, we use $^{12}$CO emission images to define a series of contour levels of increasing gas surface density, beginning with an outermost contour set to correspond to the the 3$\sigma_\mathrm{noise}$ noise level in the image that contains that GMC. Subsequent contour levels increase in steps corresponding to the $1\sigma_\mathrm{noise}$, continuing until the area enclosed by the highest surface density contour is comparable to the size of the synthesized interferometer beam, $A = \pi r_\mathrm{beam}^2$, where $r_\mathrm{beam}\approx 7$ pc at the distance of Andromeda. Within each contour $i$, we measure the enclosed area $A_i$ and equivalent radius $R_i = \sqrt{A_i/\pi}$, mass $M_i$, mean surface density $\Sigma_i = M_i/A_i$, and mean velocity dispersions in the $^{12}$CO and $^{13}$CO lines, $\sigma_{12,i}$ and $\sigma_{13,i}$.\footnote{There is one minor difference here from the procedure we use in \autoref{ssec:numerical_tests} that is worth pointing out. In the observational data we group all the area within a given contour into a single object even if that contour breaks up into two disconnected regions, whereas for the simulations we only consider connected regions. The connectedness requirement is mandatory for the simulations, because if we do not impose it then the periodic nature of the simulation domain implies that the total enclosed area is undefined, since the simulation box could be infinitely tiled. The observations, on the other hand, are of finite clouds in finite fields of view. In practice, however, whether we break up disconnected contours into multiple objects or count them as a single object has relatively little effect on the virial diagram, since splits tend to happen only at the highest contour levels; when they do occur, our choice of not breaking the contours up into multiple objects is maximally conservative when it comes to looking for upturns in the virial diagram. The reason is that, if we were to break data at a given contour level into multiple disconnected regions rather than grouping them, the main effect would be to decrease the area $A_i$ of each region, while leaving $\sigma_{12,i}$ and $\sigma_{13,i}$ mostly unchanged. This in turn would make upturns in the virial diagram at high $\Sigma$ even more prominent.} Of course $^{12}$CO emission is detected at all contour levels, but the weaker $^{13}$CO lines are measured only within those contours where the peak SNR in the $^{13}$CO line averaged over the contour is $\geq 15$. We retain only sources with at least five detected contour levels for analysis. Using this method, Lada et al. (2025, submitted) measure values of $\sigma_{12}^2/R$ for at least five surface density level $\Sigma$ for 50 individual M31 GMCs from the \citet{Lada24a} sample, and for a subset of 26 of they also measure $\sigma_{13}^2/R$ for at least five values of $\Sigma$. They use these data to determine the internal pressures within each GMC and conclude that many of the GMCs in M31 are in virial balance once the mean ISM pressure is considered. Here we use the quantities derived by the above method to additionally compute the observer's virial ratio (\autoref{eq:avir_obs}) $\alpha_{\mathrm{vir,12},i}$ and $\alpha_{\mathrm{vir,13},i}$, using the velocity dispersions in $^{12}$CO and $^{13}$CO.

\begin{figure}
    \includegraphics[width=\columnwidth]{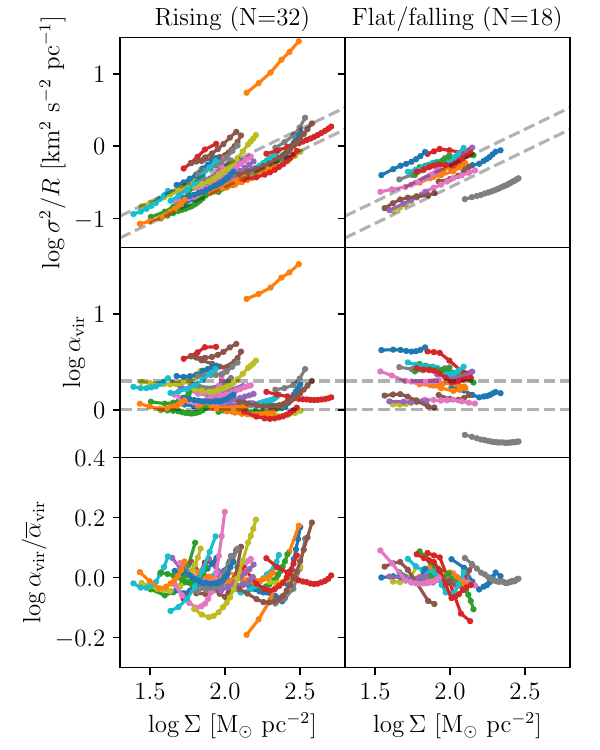}
    \caption{Differential virial analysis for GMCs in M31. The top row shows the virial diagram $\sigma^2/R$ versus $\Sigma$, with the loci corresponding to $\alpha_\mathrm{vir} = 1$ and 2 shown as the gray dashed lines. The second row shows the observer's virial parameter $\alpha_\mathrm{vir}$ as a function of $\Sigma$, while in the third row we show $\alpha_\mathrm{vir}/\overline{\alpha}_\mathrm{vir}$, where $\overline{\alpha}_\mathrm{vir}$ is the mean virial parameter computed over all contour levels; this final row is intended to focus on the shape of the $\alpha_\mathrm{vir}$ versus $\Sigma$ curve, which is the quantity of interest for differential virial analysis. Each colored curve represents a different GMC, with points along the curve indicating contour levels spaced by $1\sigma_\mathrm{noise}$ uncertainties in column density. The left column shows the ($N=32$) sources with rising virial ratios indicative of support, while the right column shows the ($N=18$) sources for which the virial ratio is flat or falling; see main text for details. \\
    \label{fig:m31_12co}}
\end{figure}

\begin{figure}
    \includegraphics[width=\columnwidth]{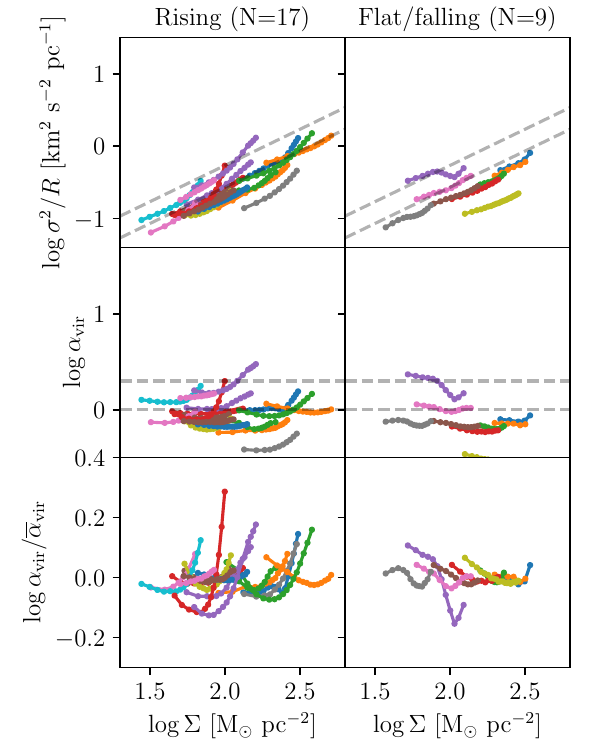}
    \caption{Same as \autoref{fig:m31_12co}, but using velocity dispersions measured from $^{13}$CO rather than $^{12}$CO.\\
    \label{fig:m31_13co}}
\end{figure}

We show our differential virial analyses using the $^{12}$CO and $^{13}$CO measurements in \autoref{fig:m31_12co} and \autoref{fig:m31_13co}, respectively. In these plots, the top rows show the traditional virial diagram, $\sigma^2/R$ versus $\Sigma$, which directly compares the independently-measured observable quantities, while the second rows shows $\alpha_\mathrm{vir,12}$ and $\alpha_\mathrm{vir,13}$, the observer's virial ratios computed using the two lines. Finally, in the bottom row we show $\alpha_\mathrm{vir}/\overline{\alpha}_\mathrm{vir}$, where $\overline{\alpha}_\mathrm{vir}$ is simply the mean of the virial ratios evaluated at each contour level we have defined for each cloud; our motivation for showing this last row is that it allows us to put all clouds on a similar footing and focus solely on the \textit{shape} of the curve traced out in the virial diagram, which is the central object in our analysis due to its much lower level of systematic uncertainty compared to the absolute value of $\alpha_\mathrm{vir}$.

For the purposes of making this plot, we have characterized shapes of the virial diagram curve as either rising, as expected for clouds that are supported, or flat or falling, as expected for free-fall collapse. To be conservative in what we call rising, we include clouds in this category only if (1) $\alpha_\mathrm{vir}$ is strictly increasing over the five highest available contour levels (i.e., if we have $N$ contour levels for a given cloud, we require that $\alpha_{\mathrm{vir},N-4} < \alpha_{\mathrm{vir},N-3} < \cdots < \alpha_{\mathrm{vir},N}$), and (2) if the virial parameter at the highest contour level  is larger than the mean virial parameter averaged over all contour levels (i.e., $\alpha_{\mathrm{vir},N} > \overline{\alpha}_\mathrm{vir}$). We perform this categorization independently for the $^{12}$CO and $^{13}$CO data sets, finding 32/50 and 17/26 rising profiles, respectively. (If we compute fraction by mass rather than number, the ratios are nearly identical.) We plot the clouds in these two categories separately in the two columns in \autoref{fig:m31_12co} and \autoref{fig:m31_13co}.

The first and most obvious conclusion to draw from these figures is that a substantial majority of the observed GMCs have rising virial ratios, and that the curves of virial ratio versus surface density that we see for these clouds look very similar to the profiles shown in, for example, the middle panel of \autoref{fig:avir_polytrope} for hydrostatic, non-singular polytropes, or the top left panel of \autoref{fig:avir_sim}, for our simulation at a time when it is not undergoing gravitational collapse. The common feature to these curves is a broad region of flat or slightly falling value of $\alpha_\mathrm{vir}$ at low $\Sigma$, following by a fairly sharp rise at the highest $\Sigma$. 

\begin{figure}
    \includegraphics[width=\columnwidth]{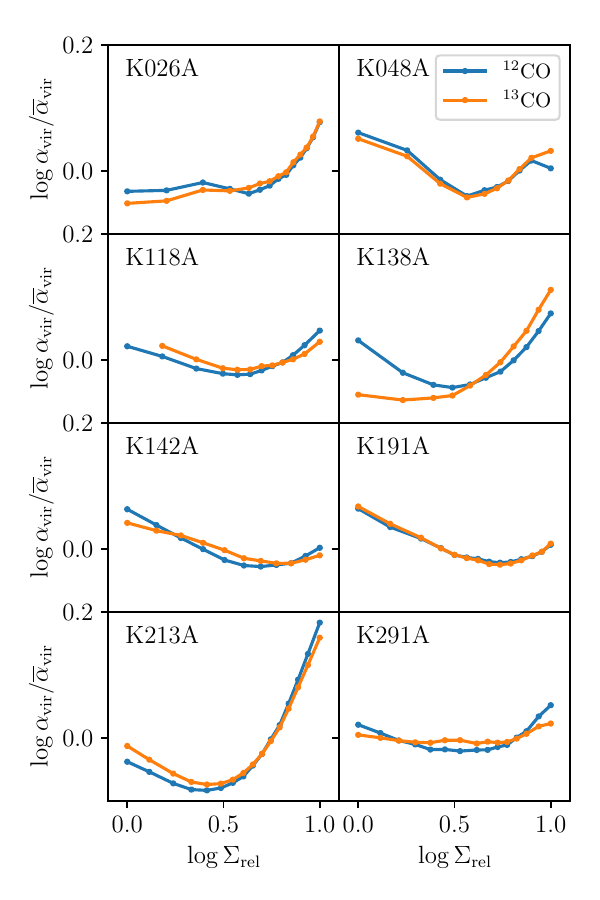}
    \caption{Comparison of differential virial analysis for velocity dispersions determined from $^{12}$CO versus $^{13}$CO data for eight randomly-selected clouds in our sample, as indicated in the labels in each panel. Vertical axes show the virial ratio normalized to the mean, $\alpha_\mathrm{vir}/\overline{\alpha}_\mathrm{vir}$, while horizontal axes show gas surface density, rescaled to run from 0 to 1 on a logarithmic scale for all clouds, i.e., for a GMC for which $\Sigma_1$ and $\Sigma_N$ are the lowest and highest contour levels available, we define $\log\Sigma_\mathrm{rel} = (\log\Sigma-\log\Sigma_1)/(\log\Sigma_N - \log\Sigma_1)$. Blue shows $^{12}$CO data, orange shows $^{13}$CO. As in \autoref{fig:m31_12co}, points show individual contour levels, separated by $1\sigma_\mathrm{noise}$ uncertainties in column density. Note that for some clouds there are fewer $^{13}$CO than $^{12}$CO data points; these correspond to clouds where $^{13}$CO was not detected with $\mathrm{SNR}>15$ at the indicated contour level.\\
    \label{fig:12_13}}
\end{figure}

A second conclusion to draw is that, even for the clouds that have flat or falling virial profiles by our definition, there are often hints of an upturn in $\alpha_\mathrm{vir}$ at the highest surface densities. These are insufficient for us to classify the clouds as rising because they are resolved by too few contour levels (for example the blue and orange curves shown in the right column of \autoref{fig:m31_13co}) or because the rise is not quite enough for $\alpha_\mathrm{vir}$ at the highest contour level to go above the mean set at lower $\Sigma$ (for example the gray and pink curves in the right column of \autoref{fig:m31_13co}), but the shape of these profiles certainly hints that they might change category from flat to rising if we were able to resolve smaller, high column density structures.

A third conclusion is that, while there are some offsets in the absolute value of $\alpha_\mathrm{vir}$ between measurements from $^{12}$CO and $^{13}$CO, the shapes of curves in the virial diagram are very similar in either tracer. This is consistent with our expectation that shapes in the virial diagram should be much more robust against systematics than absolute levels. To quantify this agreement further, in \autoref{fig:12_13} we compare $\alpha_\mathrm{vir}/\overline{\alpha}_\mathrm{vir}$ curves determined using $^{12}$CO versus $^{13}$CO data for eight randomly-selected clouds in our sample. We see that the results are extremely similar in most cases, indicating that the use of differential rather than absolute virial ratios reduces the systematic uncertainties in the analysis exactly as intended.

Based on this analysis, we are able to conclude that at least a majority of the GMCs in Andromeda are not in a state of global collapse. A minority may be, but answering that question definitively will require extending the differential virial analysis to smaller size scales in order to see whether the hints at an upturn we see at the highest surface densities are the beginnings of real features, or are random fluctuations that will flatten again.

There are two possible interpretations of this finding. One would be to interpret this as a time series, whereby clouds spend an extended period ($\gtrsim 2/3$ of their lives) in a supported state, and only begin to undergo collapse at the ends of their lives. Such extended periods of non-collapse do not appear to be compatible with most published models of global (or global hierarchical) collapse to date; for example, as discussed in \autoref{sec:intro}, \citet{Vazquez-Semadeni07a} find that the duration for which the gravitational potential energy is larger than the kinetic energy is only $\approx 1/3-1/2$ as long as the collapse phase, and \citet[their Figure 13]{Vazquez-Semadeni19a} generically predict that the collapse phase is longer than the initial mass accumulation phase in their proposed scenario. However, it seems conceivable that one could construct an alternative version of this scenario during which collapse still occurs, but only at the tail-end of much longer period of non-collapse, as the statistics of clouds in Andromeda seem to require. An alternative possibility, of course, is that the data do not represent a time sequence, and that clouds undergo brief periods (constituting $\lesssim 1/3$ of a cloud's lifetime) of large-scale collapse before being stabilized again, presumably by either feedback or an injection of fresh, turbulent material from a large-scale accretion flow (as proposed for example by \citealt{Jeffreson24a}). The data at hand do not permit us to distinguish between these possibilities.

\section{Conclusion and future prospects}
\label{sec:conclusion}

We introduce a new method, which we term differential virial analysis, for determining the dynamical state of giant molecular clouds (GMCs), and in particular for determining whether or not they are in a state of collapse. Our method is based on analysis of how clouds' virial ratios change as a function of column density or size scale. This approach has two primary advantages over traditional virial analysis. First, absolute values of virial ratios are substantially uncertain due to complications such as the unknown 3D geometry of clouds seen only in projection, but we show that the differential change in the virial ratio as a function of scale is robust against such systematics. Second, curves of virial ratio versus column density contain a distinctive diagnostic feature: in structures that are supported against collapse by internal motions or pressure, the virial ratio rises systematically as we move to smaller size and higher density scales, reflecting the existence of unbound structures that are confined by the pressure of the material around them. By contrast, clouds that are in a state of dynamical collapse have motions that are gravity-driven on all size scales, and pressure is never an important force. Consequently, they have virial ratios that remain constant as a function of size scale. Thus the existence of a ``hook'' of increasing virial ratio at high surface density is a clear marker of clouds that are not in a state of global collapse. Conversely the non-existence of such a hook could be evidence of collapse, with the caveat that one must be careful that the data are sufficiently deep -- we have shown in \autoref{sec:andromeda} there are a number of clouds that, while they do not meet our stringent criteria for the existence of a hook, show hints of one that might become clearer in deeper data.

We validate this intuitive argument by comparison to both analytic hydrostatic and collapse solutions, for which we can compute exact results, and numerical simulations that transition from supported to collapsing over time. For the latter, we show that the existence of the hook feature is a unique property of non-singular hydrostatic configurations, and that is vanishes in both singular hydrostatic and collapsing structures. With regard to the simulations, we show that the hook feature is absent at times prior to the onset of global collapse but then appears as collapse begins, demonstrating that differential virial analysis can separate these two evolutionary phases. \reply{In future work, however, it would be useful to repeat this analysis for a larger sample of simulations.}

We conclude with a discussion of systematic observational uncertainties, demonstrating that there are potential concerns but that these can be overcome with data of sufficient quality, and then providing a sample application of differential virial analysis to a sample of GMCs in Andromeda drawn from \citet{Lada24a}. We find from this analysis that a significant majority of the GMCs of Andromeda possess the distinctive feature in their differential virial diagrams that is indicative of support and inconsistent with global collapse. Moreover, many of the clouds that do not obviously show this feature do show hints at it, indicating that it might appear in higher-resolution data. We conclude therefore that most GMCs in Andromeda cannot be in a state of global collapse\reply{; this does not mean that global collapses cannot occur at all, but that, if they do, they must occupy only a small minority of the full molecular cloud life cycle}.

The technique we have developed is a general one, and can be used on any data set with enough depth to allow a significant dynamic range in column density. We therefore anticipate that it will be widely applicable to additional extragalactic GMC surveys.

\section*{Data and Software Availability Statement}

The software used to produce all the figures and carry out all the analysis presented in this paper is available on \href{https://github.com/markkrumholz/diffvirial/}{GitHub} under an open source license. This software makes use of the \textsc{NumPy} \citep{Numpy20a}, \textsc{SciPy} \citep{SciPy20a}, \textsc{AstroPy} \citep{Astropy-Collaboration13a, Astropy-Collaboration18a, Astropy-Collaboration22a}, \textsc{Scikit-image} \citep{Walt14a}, and \textsc{Matplotlib} \citep{Hunter07a} packages. The M31 GMC data required to reproduce \autoref{fig:m31_12co} through \autoref{fig:12_13} are included in the \href{https://github.com/markkrumholz/diffvirial/}{GitHub} repository, while the simulation data used to produce \autoref{fig:sim_contour} and \autoref{fig:avir_sim} are available from the \href{https://www.mhdturbulence.com/}{Catalogue of Astrophysical Turbulence Simulations (CATS)}; detailed instructions for which CATS files are required are provided in the \href{https://github.com/markkrumholz/diffvirial/}{GitHub} repository.

\section*{Acknowledgments}

This work began with a discussion between MRK and CJL at the conference ``The Fullness of Space: Celebrating the career of Christopher F.~McKee'', and the authors with to acknowledge both Chris McKee, for providing the inspiration for the conference, and Jonathan Tan, for being its chief organizer. MRK acknowledges support from the Australian Research Council through Laureate Fellowship FL220100020. We thank the Enrique V\'azquez-Semadeni and the referee, Javier Ballesteros-Paredes, for helpful comments.

\bibliographystyle{aasjournal}

\bibliography{refs}




\end{document}